# Control of excitonic absorption by thickness variation in few-layer GaSe

Arne Budweg[1], Dinesh Yadav[1], Alexander Grupp[1], Alfred Leitenstorfer[1], Maxim Trushin[2,1], Fabian Pauly[3,1,*], Daniele Brida[1,4,+]

[1] Department of Physics and Center for Applied Photonics, University of Konstanz, 78457 Konstanz, Germany

[2] Centre for Advanced 2D Materials, National University of Singapore, 6 Science Drive 2, Singapore 117546

[3] Okinawa Institute of Science and Technology Graduate University, Onna-son, Okinawa 904-0495, Japan

[4] Physics and Materials Science Research Unit, University of Luxembourg, 162a avenue de la Faïencerie, 1511 Luxembourg, Luxembourg

* email: fabian.pauly@oist.jp

+ email: daniele.brida@uni.lu

ABSTRACT

We control the thickness of GaSe on the level of individual layers and study the corresponding optical absorption via highly sensitive differential transmission measurements. Suppression of excitonic transitions is observed when the number of layers is smaller than a critical value of 8. Through ab-initio modelling we are able to link this behavior to a fundamental change in the band structure that leads to the formation of a valence band shaped as an inverted Mexican hat in thin GaSe. The thickness-controlled modulation of the optical properties provides attractive resources for the development of functional optoelectronic devices based on a single material.

INTRODUCTION

Recent investigations of the electronic structure of layered materials have led to disruptive advances in photonics, optoelectronics and spintronics [1-6]. In this context, metal monochalcogenides form one subgroup with similar structural properties that allow thickness reduction down to atomically thin two-dimensional (2D) crystals with a great variety of unique qualities. GaSe is a prominent member of this group that is widely used in near- and mid-infrared nonlinear optics [7,8]. First studies using few-layer GaSe in transistors [9] and photodetectors [10,11] have been performed. Recent photoluminescence measurements on GaSe nanosheets demonstrated a high preservation of photon polarization between absorption and photoluminescence, indicating a high potential for spintronic applications [12,13]. Furthermore, theoretical calculations suggest carrier-induced magnetism in single-layer GaSe [14].

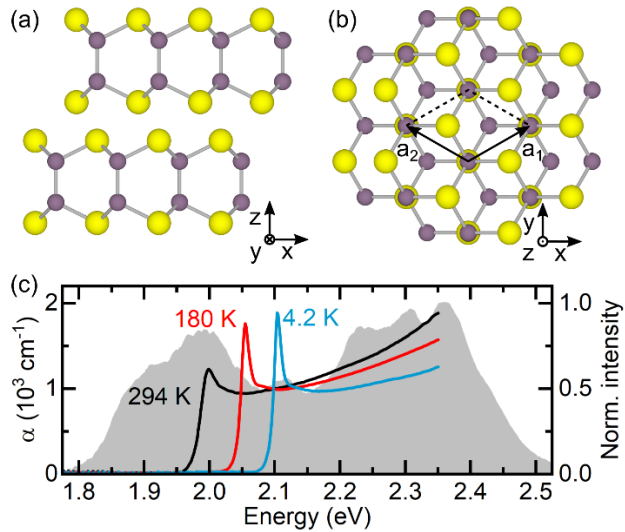

Fig 1: Schematic crystal structure of ε-GaSe viewed perpendicularly to the (a) *y*- and (b) *z*-axis. Yellow and violet spheres represent selenium and gallium atoms, respectively. (c) Measured absorption coefficient for bulk ε-GaSe at different temperatures (solid lines) and normalized

intensity spectrum of the pump pulses (gray area) employed in the transient absorption experiments.

In this work, we demonstrate experimentally the possibility to control the excitonic absorption from substantial enhancement to complete suppression by changing the number of GaSe layers in a nanosheet. Theoretical modelling tentatively explains this behavior as follows: The electron-hole attraction at a thickness of approximately 9 nm is competing with the inversion of the valence band curvature at the same thickness, leading to a strongly reduced light-matter coupling in few-layer GaSe.

RESULTS AND DISCUSSION

GaSe is a III-VI semiconductor, forming crystals with a hexagonal layered structure. Individual layers consist of four atomic planes in a Se-Ga-Ga-Se order and interlayer bonding arises from weak van der Waals forces. In the ε-GaSe polytype the stacking follows an AB sequence, corresponding to the non-centrosymmetric space group $D_{3h}^1$ [15,16]. Thickness values between 0.8 and 1 nm are reported for a GaSe monolayer [13,17,18]. Figure 1 displays the crystal structure of a double layer of GaSe in (a) side and (b) top view, stacked according to the ε polytype. At room temperature, bulk GaSe exhibits a quasi-direct bandgap of approximately 2 eV. The valence band maximum (VBM) is at the Γ point, while the global conduction band minimum (CBM) is positioned in proximity of the M point. However an almost isoenergetic local minimum exists at the Γ point, exceeding the M valley by only 10 meV [19]. The direct exciton with a binding energy of 19 meV is also located at the Γ point, in close energetic vicinity to the direct and indirect interband transitions [15,20].

All samples are fabricated by mechanical exfoliation of a Bridgman grown ε-GaSe single crystal. This material is first thinned down to a thickness of about 1 μm by successive exfoliation with adhesive tape. The flakes are then transferred to a 150 μm thick $SiO_2$ substrate. By continuing the exfoliation on the substrate, we produce large homogenous areas of few-layer GaSe with thicknesses down to 4 nm. Residual adhesive is removed by cleaning the surface with acetone and isopropanol. Some samples are additionally sputter-coated with a 15 nm protective layer of $Si_3N_4$ to exclude oxidation of GaSe at ambient conditions. All specimens are stored under nitrogen atmosphere.

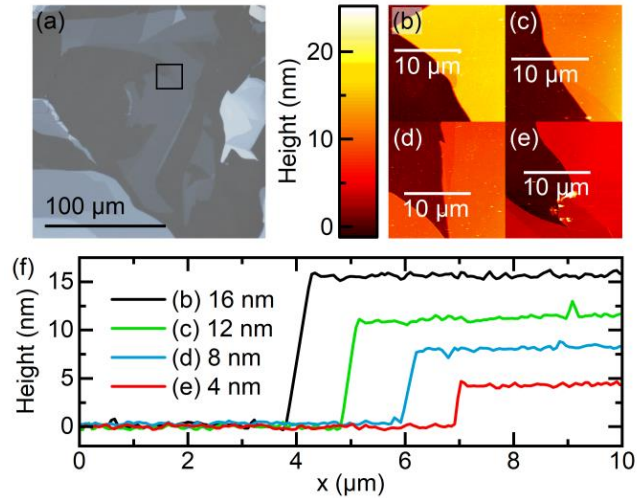

Fig 2: (a) Optical micrograph of few-layer GaSe on $SiO_2$. Height profiles extracted from topographical maps along the white lines in panels (b)-(e) are depicted in panel (f), revealing thicknesses between 4 and 16 nm. The black box in panel (a) indicates the area corresponding to the data reported in panel (e).

Few-layer flakes are identified in an optical microscope, as depicted in Figure 2(a). In addition, topographical characterization of the relevant areas on the sample is performed with an atomic force microscope (AFM) operating in tapping mode. Figures 2(b)–2(e) show AFM scans of four different regions with thicknesses ranging between 4 and 16 nm. In all maps, the white scale bars indicate a length of 10 µm. Height profiles along these lines are plotted in Fig. 2(f). Adopting a single-layer thickness between 0.8 and 1 nm [13,17,18], these values correspond to a number of layers between 4 and 20 for the four distinct specimens.

Transient absorption measurements of the samples are performed with an ultrafast laser system operating at a repetition rate of 50 kHz [21]. A custom-designed noncollinear optical parametric amplifier delivers pump pulses with a bandwidth of 0.52 eV at a central energy of 2.15 eV, as plotted in Fig. 1(c). These pulses are compressed to a temporal duration of 7 fs by means of customized dielectric chirped mirrors (DCMs). The pump-induced transmission change is probed by a white-light supercontinuum comprising spectral components between 1.7 and 2.4 eV at a pulse duration of 10 fs. Pump and probe pulses are focused onto the sample under a small angle to separate the beams after transmission through the sample. Using an off-axis parabolic mirror with a focal length of 50.8 mm, we reach focus diameters of 30 µm for the pump and 15 µm for the probe beam while keeping the excitation density at a moderate value of 200 µJ/cm$^2$. Orthogonal polarization between the beams enables us to filter out residual pump light before detection. The experiments are performed in a cryostat with the possibility to control the sample temperature down to 4.2 K. The probe light, collected after interaction with the sample, is spectrally dispersed and then detected by a fast CCD linear array [21] for the measurement of the differential transmission.

We have exploited the same broadband probe beam to also determine the linear absorption spectra of a 25 µm thick free-standing GaSe crystal mounted inside the cryostat, which we consider as a model for the bulk limit. The solid lines in Fig. 1(c) depict the measured absorption coefficient for this GaSe sample at temperatures of 294 K, 180 K and 4.2 K. The curves show a pronounced shift of the absorption maximum corresponding to the direct excitonic transition and significant line broadening as a function of the sample temperature.

As one of the most fundamental characterization tools, linear absorption provides valuable insight into the optical properties of a material, emphasizing especially the transitions carrying a strong dipole moment. Nevertheless, it is challenging to study with a high sensitivity the very small absorption of few-layer GaSe that is estimated to be in the order of 0.01%. To monitor the changes in the optical properties of few-layer GaSe, we therefore employ transient absorption spectroscopy. This technique provides high-frequency modulation capabilities, thus enabling us to detect precisely relative changes in the probe transmission with a sensitivity level better than $10^{-5}$. For this reason, we perform degenerate pump-probe experiments, where the pump excites electronic transitions up to energies well above the bandgap and the probe detects the changes occurring at photon energies overlapping with the bandgap and related excitonic resonances. Figure 3 shows the spectrally resolved temporal dynamics in a several hundred nm thick flake of GaSe at a temperature of 4.2 K as a bulk reference. The map of relative transmission change exhibits a distinct derivative-like feature between 2.07 and 2.16 eV which can be attributed to the excitonic resonance of GaSe. After a quasi-instantaneous onset of the signal, the thermalization of the initial out-of-equilibrium charge carrier distribution occurs within the first picosecond and it is followed by a significantly slower recombination process on a timescale of several hundred picoseconds. The differential transmission signal is induced by the bandgap renormalization that causes a broadening and a shift of the excitonic transition. In addition, the photoexcitation of carriers reduces the probability for further probe photon absorption with a consequent bleaching signature that competes with the bandgap renormalization [22,23].

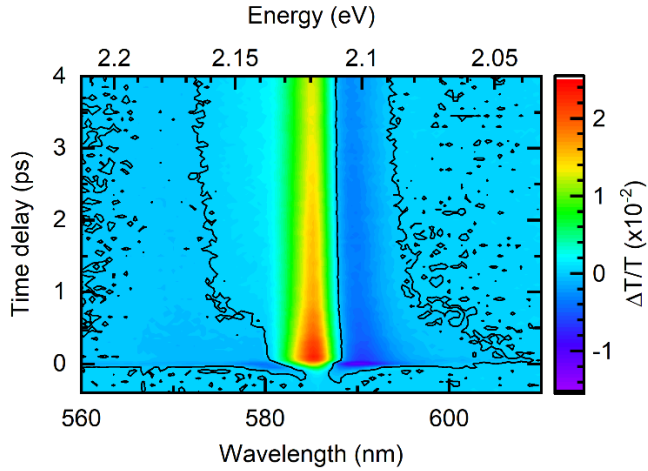

Fig 3: Differential transmission change of bulk GaSe at 4.2 K as a function of pump-probe time delay and probe wavelength.

In Fig. 4 the differential transmission spectrum at a fixed temporal delay between pump and probe pulse is plotted for several different few-layer GaSe samples. These time delays are selected to identify the response of the samples after the first ultrafast thermalization has occurred. Panels (a) and (b) of Fig. 4 display datasets acquired at temperatures of 180 K and 4.2 K, respectively, to assess the effect of line broadening. It appears that renormalization is dominant at low temperature with a derivative-like differential spectrum. Instead, bleaching is more significant at 180 K since at this temperature the excitonic line is already broadened and the transient transmission signal appears as enhanced transmission of the probe spectrum, i.e. positive $\Delta T/T$.

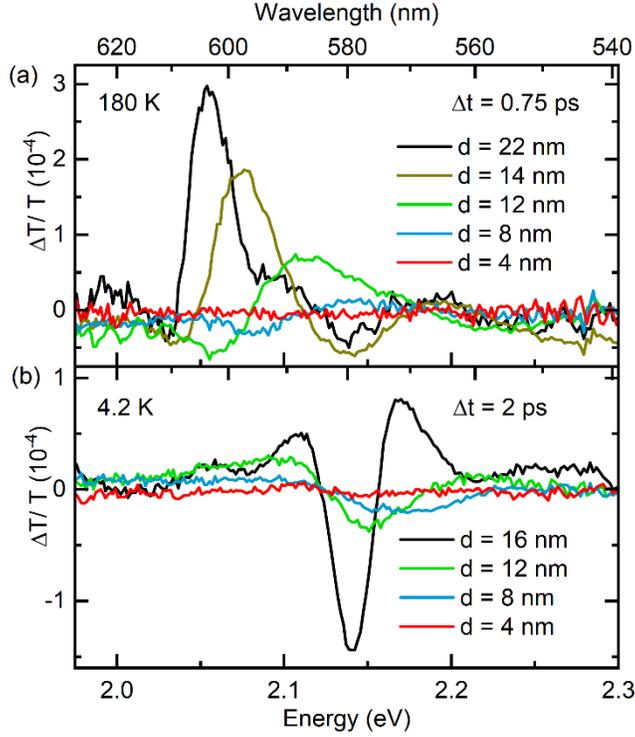

Fig 4: Relative transmission change Δ*T*/*T* after optical excitation of few-layer GaSe at (a) 180 K and (b) 4.2 K for a set of samples with different thickness *d*.

Independently of the spectral shape of the differential transmission, we observe in Fig. 4 a broadening and blue-shift of the excitonic signature as well as a reduction in its strength by decreasing the number of GaSe layers. Remarkably, the evolution of the signal is rather abrupt for samples thinner than 15 nm, with a two-fold reduction of its amplitude when reducing the sample from 14 nm to 12 nm at 180 K and five-fold between 16 nm and 12 nm at 4.2 K. For thicknesses below 8 nm the differential transmission change is no longer detectable even with our sensitive method, indicating the suppression of excitonic absorption. This behavior is confirmed by further pump-probe measurements at different temperatures and as a function of the GaSe sample thickness. Fig. 5(a) depicts the blue-shift of the photon energy associated with the exciton transition as extracted from these experiments. In detail the values are extracted from the pump-probe spectra by tracking the probe wavelength associated with the maximum signal in the differential transmission. Fig. 5(b) shows the magnitude of the excitonic signal as a function of the GaSe sample thickness. The dashed lines represent linear fits of the data points for thickness values up to 16 nm. It is clear that the excitonic signature vanishes for flakes thinner than 8 nm. Importantly, a significant residual signal cannot be hidden under the technical noise of the experiments, since the limit of the sensitivity is $10^{-5}$ (dotted black line in figure 5(b)).

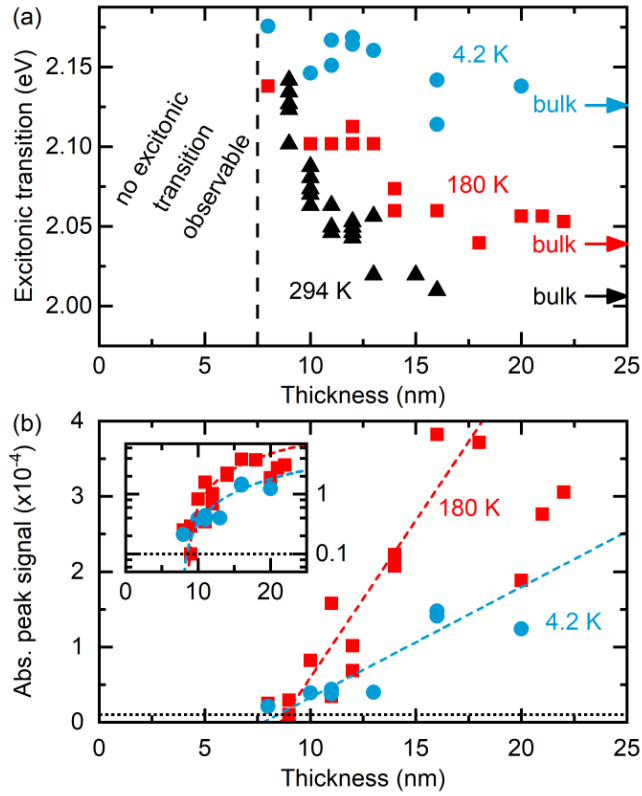

Fig 5: Measured energy position (a) and signal strength (b) of the maximum transmission change after optical excitation for different GaSe areas with thicknesses ranging from 8 to 25 nm and at temperatures of 4.2 K (blue circles), 180 K (red squares) and 294 K (black triangles). The arrows in panel (a) mark the energy of the exciton transition in bulk samples for each temperature. In panel (b), the dashed lines represent linear fits of the data points while the inset shows the same information on a logarithmic scale. The dotted line represents the noise level of the experiments.

A similar reduction in signal strength, as observed in our data, was recently reported for the photoluminescence of GaSe and attributed to rapid oxidation of thin flakes [24]. In contrast, we prevent oxidation of some samples by passivating them with a layer of $Si_3N_4$ right after exfoliation. Both sample sets, with and without passivation, display the vanishing signal at excitonic transitions, thus excluding oxidation as its origin. Instead, we tentatively attribute the observed behavior to a fundamental shaping of the valence band structure [25] in few-layer GaSe combined with substrate screening, as explained in the following.

The band structure of single-layer GaSe has been calculated and studied thoroughly [14,25-28], and band calculations for few-layers have been performed both for the β-GaSe polytype [25] and the ε polytype [29]. For the ε-GaSe polytype, which is relevant for exfoliated samples [30], we perform ground-state electronic structure calculations using density functional theory (DFT) within the local density approximation (LDA) [31], as implemented in the open-source package QUANTUM ESPRESSO [32]. We employ a plane-wave basis set with a kinetic energy cutoff of 80 Ry and a charge-density cutoff of 320 Ry together with fully relativistic norm-conserving pseudopotentials [33]. For the bulk we use a Γ-centered 16×16×3 k-grid. Few-layer structures are calculated by keeping the distance between periodic images constant at 18 Å along the stacking direction [z-axis in Fig. 1(a)] and the Brillouin zones are sampled with a 12×12×1 k-grid centered at the Γ point. Geometries are subsequently optimized by using the Broyden-Fletcher-Goldfarb-Shanno algorithm until the net force on atoms is less than $10^{-4}$ $N$ Ry/a.u. and the total energy change is below $10^{-6}$ $N$ Ry, where $N$ is the number of GaSe layers in our supercell. We consider spin-orbit coupling (SOC) in the band structure calculations but omit it in geometry optimizations, since it does not play a major role in that step.

The evolution of the highest valence band (HVB) and the lowest conduction band (LCB) is shown in Fig. 6(a). Since splitting of the bands around the Γ point is small, only one spinor-component is displayed. Similar to previous reports [25,29], our calculations show that the HVB evolves from a parabolic shape to an inverted Mexican hat by reducing the number of layers. In this process, the effective mass of the holes changes from positive to negative at the Γ point. This transition occurs at the critical value of 8 layers, where the HVB is approximately flat around the center of the Brillouin zone. In contrast to the HVB, the curvature of the LCB is qualitatively unaffected by the GaSe thickness, exhibiting a parabolic shape equivalent to a positive effective mass of the electrons. As shown in Fig. 6(b), the gap between the LCB and HVB at the Γ point increases with decreasing layer thickness, as expected from quantum confinement effects due to dimensionality reduction.

We investigate further the absorption spectra of the layers by computing the complex dielectric function $\varepsilon(E) = \varepsilon_1(E) + i\varepsilon_2(E)$ within the random phase approximation (RPA) using QUANTUM ESPRESSO [32]. While we neglect excitonic effects in this approach, we concentrate on vertical transitions within the single-particle band structure, leading to non-interacting electron-hole pairs. We find that without SOC transitions from the HVB to the LCB at the Γ point are strictly forbidden in all structures of the ε-GaSe polytype independent of layer thickness, if the light is polarized in-plane. Selection rules are changed, however, with the inclusion of SOC, making such transitions weakly allowed, as has been pointed out for the bulk and single-layer cases [15,34]. We note that stacks with even layer numbers possess point-group symmetry $C_{3v}$, while those with an odd number are of $D_{3h}$ symmetry. In the RPA calculations we use a 50×50×1 $k$-grid to obtain the in-plane (straight lines) and out-of-plane (dashed lines) components of the imaginary part of the dielectric function $\varepsilon_2(E)$, as depicted in Fig. 6(c), which is proportional to the absorption coefficient. The absorption spectra are normalized to a single layer by plotting $\varepsilon_2^{(1)}(E) = L\epsilon_2(E)/L_s$, where $L_s$ is the thickness of the GaSe sample and $L$ is the total thickness of the supercell perpendicular to the stack, thus also including the surrounding vacuum.

Fig. 6(c) shows that, within our assumption of non-interacting electrons and holes, the absorption of the individual layers does not change significantly as the number of layers in the GaSe crystal is reduced and that it is stronger if the light is polarized out-of-plane. However, the strongly focused beams employed in the experiments access both in- and out-of-plane components of the absorption spectrum with negligible dependence of the differential transmission signal on the angle of incidence onto the samples.

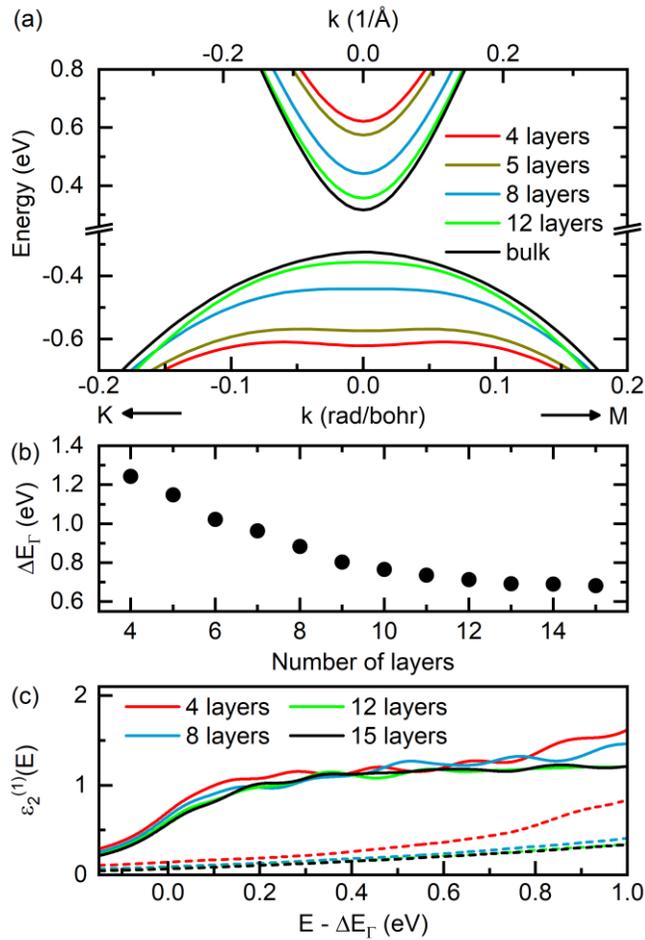

Fig 6: (a) Evolution of HVB and LCB with increasing number of layers for the ε-GaSe polytype. (b) Change of the DFT bandgap at the $\Gamma$ point. (c) Absorption spectra of 4, 8, 12 and 15 layers calculated within RPA as a function of the photon energy. The spectra are plotted relative to the bandgap at the $\Gamma$ point and are normalized to a single layer, as described in the text. Solid lines show $\varepsilon_2^{(1)}(E)$, if the light is polarized out-of-plane [$\varepsilon_{2,zz}^{(1)}(E)$], dashed lines are for the in-plane direction [$\varepsilon_{2,xx}^{(1)}(E) = \varepsilon_{2,yy}^{(1)}(E)$, see Fig. 1].

Since qualitative changes of the absorption based on band structure effects can be excluded in a single-particle picture [see Fig. 6(c)], we now focus on the relation between the slab thickness $d$ and many-body interactions as characterized by the ground-state exciton radius $R_0$. Assuming an anisotropic hydrogenic model for Wannier-Mott excitons [35], we can estimate the exciton radius as $R_0 = e^2/(2\sqrt{\varepsilon_\parallel \varepsilon_\perp} E_0)$. Using literature values [36] for the binding energy $E_0 = 19.7$ meV as well as for out-of-plane and in-plane dielectric constants $\varepsilon_\parallel = 6.18$ and $\varepsilon_\perp = 10.6$, respectively, we find $R_0 = 4.5$ nm in the bulk limit. Surprisingly, the corresponding exciton diameter of 9 nm is very close to the critical slab thickness, below which the excitonic signal disappears in the optical excitation measured [see Fig. 5(b)]. Once $d \leq 2R_0$ the 3D exciton does not fit the out-of-plane dimension anymore and hence must be considered as a 2D entity. Apparently, this critical thickness corresponds to just above 8 layers, where the valence band acquires the inverted Mexican hat shape with the in-plane wave vector of the local band maxima $k_0 \approx 0.1$ Å$^{-1}$ [see Fig. 6(a)]. Thus, the critical thickness of about 9 nm (i) leads to the suppression of the exciton absorption line, (ii) induces a valence band inversion, and (iii) separates the 2D and 3D regimes in the electron-hole relative motion. This suggests that phenomena (i)–(iii) should be interrelated.

The valence band inversion may indeed lead to qualitative changes in the electron-hole relative motion and exciton ground state. The radial wave vectors relevant for the excitonic ground state formation lie below $1/R_0$. Since $k_0 R_0 > 1$ in our case, both the valence and conduction bands are characterized by a positive curvature at the relevant momentum scale, yielding a configuration where both conduction and valence band states have positive group velocities. This contrasts with the GaSe monolayer limit in vacuum [28], where the binding energy is predicted to be about 660 meV, suggesting an excitonic radius at least one order of magnitude smaller than $R_0$ estimated from bulk parameters. Resulting in $k_0 R_0 < 1$, this condition allows for the existence of conventional but tightly bound excitonic states [37], since the valence band acquires the usual negative curvature at the scale of $1/R_0$, leading to a negative group velocity for the corresponding states. In real samples like ours this effect is however limited by the dielectric screening of the substrate.

The energy shift of the main excitonic feature in Fig. 5 is due to the increasing energy gap between the CBM and VBM as is qualitatively confirmed by the values calculated in Fig. 6(b), where the typical underestimation of the DFT bandgap has to be taken into account [38]. Remarkably, the excitonic transitions depicted in Fig. 4 show that the blue-shift is accompanied by a broadening of the line shapes occurring between 12 and 8 layers, possibly pointing to a weakening of the exciton transition towards its suppression in even thinner flakes.

CONCLUSION

By performing highly sensitive differential transmission measurements in GaSe, we followed the change of the optical absorption from bulk to the 2D case and observed the suppression of excitonic transitions when the number of layers is smaller than the critical value of 8. We suggest that this behavior can be assigned to a fundamental change of the valence band shape in thin

GaSe, which is confirmed by our ab-initio modelling that however does not conclusively explain if the exciton is suppressed or becomes weakly coupled to dipole excitation. In any case, our results indicate that the modulation of the optical properties of GaSe arises from an interplay of both intrinsic and extrinsic effects due to substrate screening. The discovery of this mechanism paves the way towards devices employing a single material with engineered band topology in a controlled environment, where the layer number determines the properties of light-matter interaction and complex adjacent functionalities can be integrated on a single chip.

ACKNOWLEGEMENTS

D.B. acknowledges financial support from the Emmy Noether program (BR 5030/1-1) of the German Research Foundation (DFG). D.Y. and F.P. thank the Carl Zeiss Foundation as well as the Collaborative Research Center (SFB) 767 of the DFG for funding. M.T. is supported by the Director's Senior Research Fellowship from the Centre for Advanced 2D Materials at the National University of Singapore (NRF Medium Sized Centre Programme R-723-000-001-281). Part of the numerical modeling was performed using the computational resources of the bwHPC program, namely the bwUniCluster and the JUSTUS HPC facility.

19